\documentclass[10pt]{article}

\begin{document}

\title{Testing Dark Energy  and Cardassian Expansion for Causality}
\author{J. Ponce de Leon\thanks{E-Mail:
jponce@upracd.upr.clu.edu, jpdel@astro.uwaterloo.ca}  \\Laboratory of Theoretical Physics, 
Department of Physics\\ 
University of Puerto Rico, P.O. Box 23343,  
Rio Piedras,\\ PR 00931, USA}
\date{December 6, 2004}

\maketitle
\begin{abstract}
Causality principle is a powerful criterion that allows us to discriminate between what is possible or not. In this paper we study the transition from decelerated to accelerated expansion in the context of Cardassian and dark energy models. We distinguish two important events during the transition. The first one  is the end of the matter-dominated phase, which occurs at some time $t_{eq}$. The second one  is the actual crossover from deceleration to acceleration, which occurs at some $t_{T}$. Causality requires $t_{T} \geq t_{eq}$. 
We demonstrate that dark energy models, with constant $w$, and Cardassian  expansion, are compatible with causality only if  $(\Omega_{M} - \bar{q}) \leq 1/2$. However, observational data indicate that the most probable option is $(\Omega_{M} - \bar{q}) > 1/2$. Consequently, the transition from deceleration to acceleration in dark energy and Cardassian models occurs before the matter-dominated epoch comes to an end, i.e., $t_{eq} > t_{T}$. Which contradicts causality principle.

\end{abstract}

\section{Introduction}

Currently,  there is a general agreement among cosmologists that: (i) the universe is speeding up, instead of slowing down; (ii) the accelerated expansion is a recent phenomenon; (iii) the universe is spatially flat; (iv) ordinary matter in the universe, including dark matter,  can only account for $30\%$ of the critical density. Evidence  in favor of these results is provided by observations of high-redshift supernovae Ia \cite{Riess}-\cite{Tonry}, as well as  other observations, including  the cosmic microwave background and galaxy power  spectra \cite{Lee}-\cite{Sievers}.

Since the gravity of both matter and radiation is attractive, the accelerated expansion requires 
either modified Einstein equations or, in the context of general relativity, the presence  of a mysterious form of matter, which accounts for  $70\%$ of the total content of the universe and  remains unclustered on all scales where gravitational clustering of ordinary matter  is seen.  Well-known theories that illustrate these alternatives are Cardassian expansion and dark energy models, respectively.

The universe was decelerating and dominated by matter and radiation in the past. The energy densities driving the current accelerated expansion  became dominant only recently. 
In this paper, we distinguish to important events in the process of transition from deceleration to acceleration. The first event is the end of the matter-dominated phase, which occurs at some time $t_{eq}$. The second event  is the actual crossover from deceleration to acceleration, which occurs at some $t_{T}$.
Since the transition to an accelerated epoch is a consequence of the domination of the repulsive component, it follows that there is a {\it causal connection} between these two times; namely, $t_{T} \geq t_{eq}$. In terms of the redshift, this is equivalent  to $z_{T} \leq z_{eq}$ because $z$ decreases monotonically with time.

The purpose of this paper is to test dark energy and Cardassian models for causality. We show that, in these models, the transition to an accelerated phase  is not causal. Namely, we calculate $z_{eq}$ and $z_{T}$ (or equivalently $t_{eq}$ and $t_{T}$) and find that $z_{eq} < z_{T}$ ($t_{T} < t_{eq}$), which means the transition to an accelerated phase occurs during the matter-dominated phase, {\it long before} the  source of acceleration starts to dominate over ordinary matter.

\section{End of the matter-dominated phase}

In this section we write the equations of Cardassian and dark energy models in terms of the parameter $z_{eq}$, which marks the end of the matter-dominated phase. We calculate the acceleration of the universe at the end of this phase in terms density and deceleration parameters.

\paragraph{Cardassian expansion:} In Cardassian models the FRW  equation is modified from its usual form,  $H^2 = (8\pi G/3)\rho$,  to
\begin{equation}
\label{Hubble parameter}
H^2 = A \rho + B \rho^n,
\end{equation}
where $A$, $B$ and $n$ are constants. The second term causes accelerated cosmic expansion at late times if 
\begin{equation}
\label{n}
n < 2/3.
\end{equation}
The {\it auxiliary} parameter $z_{eq}$ is the redshift at which the two terms in the r.h.s. of (\ref{Hubble parameter}) become equal to each other, viz., $A\rho(z_{eq}) = B\rho^n(z_{eq})$. Thus,
\begin{equation}
\label{B}
B = A [\rho(z_{eq})]^{(1 - n)}.
\end{equation}
Since $(n - 1) < 0$, it follows that $A\rho >  B\rho^n $ for  $z > z_{eq}$ and  $A\rho <  B\rho^n $ for  $z < z_{eq}$. We impose no restrictions on the parameter $z_{eq}$; we obtain its value from our study here. 

Substituting (\ref{B}) into (\ref{Hubble parameter}) and evaluating today, we get
\begin{equation}
\label{A}
A = \frac{{\bar{H}}^2}{\bar{\rho}\left[1 + \rho^{(1 - n)}(z_{eq}){\bar{\rho}}^{(n - 1)}\right]},
\end{equation}
where $\bar{H}$ and $\bar{\rho}$ represent the current values of the Hubble constant and matter density, respectively. Now, using (\ref{B}) and (\ref{A}) in (\ref{Hubble parameter}), we find
\begin{equation}
\label{Hubble parameter 2}
H^2 = \frac{{\bar{H}}^2}{F}\left(\frac{\rho}{\bar{\rho}}\right)\left[1 +  (F - 1)\left(\frac{\rho}{\bar{\rho}}\right)^{(n - 1)}\right],
\end{equation}
where 
\begin{equation}
\label{F}
F = 1 + \rho^{(1 - n)}(z_{eq}){\bar{\rho}}^{(n - 1)}.
\end{equation}
In Cardassian models, by assumption,   there is no vacuum contribution:  only radiation and ordinary matter\footnote{Ordinary matter includes baryonic and non-baryonic dark matter.} contribute to the expansion of the universe. Thus, in units of the critical density, we have\footnote{Here $\Omega_{M}$, $\Omega_{R}$, $\Omega_{B}$ and $\Omega_{WIMP}$ denote the density parameters of ordinary matter, radiation, baryonic matter and non-baryonic dark matter, respectively.}   $\Omega_{Total} = \Omega_{M} + \Omega_{R}$, with $\Omega_{M} = \Omega_{B} + \Omega_{WIMP}$, 
Current observations suggest $\Omega_{WIMP} \approx 0.35$, $\Omega_{B} \approx 0.02h^{- 2}$ and $\Omega_{R} \approx 2.5 \times 10^{- 5}h^{- 2}$, where $h$ is the Hubble constant today in units of $100$ km/s/Mpc. Thus, $\Omega_{B} \approx 800 \Omega_{R}$. Neglecting radiation we have $\rho = \rho_{M} = \bar{\rho}(1 + z)^3$. Therefore, in Cardassian models  the Friedmann equation for the expansion  rate becomes
\begin{equation}
\label{Friedmann eq. in Cardasian models}
H^2 = \frac{{\bar{H}}^2}{F}\left[(1 + z)^3 + (F - 1)(1 + z)^{3n}\right],
\end{equation}
where $F$ now is
\begin{equation}
\label{F for ordinary matter}
F(z_{eq}, n) = \frac{[1 + (1 + z_{eq})^{3(n - 1)}]}{(1 + z_{eq})^{3(n - 1)}}.
\end{equation}
So far there is no unique explanation for the origin of the term $B\rho^n$ in (\ref{Hubble parameter}). It may arise either as a consequence of embedding our universe as a brane in higher dimensions, or from some (yet unknown) modified Einstein's equation.

\paragraph{Dark energy:} Within the context of four-dimensional general relativity, the source of cosmic acceleration is usually called {\it dark energy}. The simplest candidate for dark energy is the cosmological constant $\Lambda$ \cite{Peebles}-\cite{Padmanabhan0}. In this approach, $\Lambda$ is introduced  by ``hand"  as a parameter in Einstein's theory of
gravity.  However, if $\Lambda$ remains constant one faces the problem of fine tuning or ``cosmic coincidence problem" \cite{Zlatev}, which refers to the coincidence that $\rho_{\Lambda}$ and $\rho_{M}$ are of the same order of magnitude today.

A phenomenological solution to this problem is to consider a time dependent cosmological term, or an evolving scalar field known as {\it quintessence} \cite{Zlatev}-\cite{Deustua}. 
In these models, the dark energy component can be considered to be a smooth fluid characterized by the equation of state $p_{D} = w \rho_{D}$, where $D$ stands for dark energy and $w$ may vary with time. If the dark energy is the vacuum energy, i.e. $\Lambda$, then $w = -1$.

Neglecting radiation, the Friedmann equation in dark energy models with constant  $w$  is given by
\begin{equation}
\label{Friedmann equation in dark energy models}
H^2 = {\bar{H}}^2\left[\Omega_{M}(1 + z)^3 + \Omega_{D}(1 + z)^{3(1 + w)}\right],
\end{equation} 
where   
\begin{equation}
\label{w}
- 1 \leq w < - 1/3.
\end{equation}
In this range, dark energy violates the strong energy condition, but satisfies the dominant energy condition\footnote{The case $w < -1$, which corresponds to what is called Phantom fields, violate both energy conditions.} 

In terms of $z_{eq}$, the density parameters $\Omega_{M}$ and $\Omega_{D}$,   can be written as
\begin{equation}
\label{Density parameters}
\Omega_{M} = \frac{(1 + z_{eq})^{3w}}{1 + (1 + z_{eq})^{3w}}, \;\;\; \Omega_{D} = \frac{1}{1 + (1 + z_{eq})^{3w}}.
\end{equation}
Therefore, the  evolution equation (\ref{Friedmann equation in dark energy models}) becomes 
\begin{equation}
\label{Friedmann equation in dark energy models in terms of zeq}
H^2 = {\bar{H}}^2\frac{(1 + z_{eq})^{3w}}{1 + (1 + z_{eq})^{3w}}\left[(1 + z)^3 + \frac{(1 + z)^{3(1 + w)}}{(1 + z_{eq})^{3w}}\right].
\end{equation}
If we compare (\ref{Friedmann eq. in Cardasian models})-(\ref{F for ordinary matter})  with (\ref{Density parameters})-(\ref{Friedmann equation in dark energy models in terms of zeq}) we see that Cardassian and quintessence models are {\it mathematically} equivalent to each other. The following identification connects the two models
\begin{equation}
\label{correspondence}
n = 1 + w,\;\;\;\Omega_{M} = \frac{1}{F}, \;\;\;\Omega_{D} = \frac{F - 1}{F},\;\;\;F(z_{eq}, w) = \frac{[1 + (1 + z_{eq})^{3w}]}{(1 + z_{eq})^{3w}}.
\end{equation}
\paragraph{Deceleration parameter at the end of matter-dominated phase:}
The deceleration parameter, $q = \ddot{a}a/\dot{a}^2$,  in  both Cardassian and  quintessence models is given by 
\begin{equation}
\label{deceleration parameter}
q(z, z_{eq}, w) = \frac{[1 + (F - 1)(3w + 1)(1 + z)^{3w}]}{2[1 + (F - 1)(1 + z)^{3w}]}.
\end{equation}
Evaluating this expression at $z_{eq}$ we obtain $q_{z_{eq}}$, the deceleration at the end of the matter-dominated phase. Using (\ref{correspondence})  and (\ref{deceleration parameter}), we find
\begin{equation}
\label{acc at zeq}
q_{z_{eq}}(w) = \frac{2 + 3 w}{4}.
\end{equation}
Now, the current cosmic acceleration $\bar{q}$ is obtained from (\ref{deceleration parameter}) evaluated at $z = 0$. Since $F = 1/\Omega_{M}$, we get\footnote{We note that accelerated expansion $\bar{q} < 0$ requires
$w < (- 0.370, \;\;-0.417, \;\;- 0.476)$, for $\Omega_{M} = (0.1,\;\;0.2,\;\;0.3)$,
respectively.}

\begin{equation}
\label{q today in terms of Omega and w}
\bar{q}(\Omega_{M}, w) = \frac{3w(1 - \Omega_{M}) + 1}{2}.
\end{equation}
Consequently, $q_{z_{eq}}$ can be written as
\begin{equation}
\label{acc at the end of matter dominated phase}
q_{z_{eq}}(\Omega_{M}, \bar{q}) = \frac{1 - 2(\Omega_{M} - \bar{q})}{4(1 - \Omega_{M})}.
\end{equation}
Thus, knowing the values of $\Omega_{M}$ and $q$ today we can predict the cosmic acceleration at the end of the matter-dominated phase.

\section{Causality}
Let us now study 
causality in these models. First, the redshift of transition  from deceleration to acceleration, which we denote as $z_{T}$, is the solution of $q(z_{T}) = 0$. From  (\ref{deceleration parameter}) we get
\begin{equation}
\label{causality condition on w}
\frac{1 + z_{T}}{1 + z_{eq}} = f(w),
\end{equation}
with
\begin{equation}
\label{f(w)}
f(w) = \left(\frac{- 1}{1 + 3w}\right)^{1/3w}.
\end{equation}
Certainly, $f(w)$ must be positive for all values of $z$ and $z_{eq}$. This requires  $w < - 1/3$, which is compatible with the condition $n < 2/3$ for acceleration in Cardassian models (\ref{n}). On the other hand, for causality reasons\footnote{Since $dz/dt = - (1 + z)H$, and $H > 0$,  it follows that $z$ decreases {\it monotonically} with time. In other words, if $t_{2} > t_{1}$, then $z_{2} < z_{1}$.}, it is clear that the timing of $z_{eq}$ must be earlier (or not later than) the one for $z_{T}$.   In other words, causality requires $z_{eq} \geq z_{T}$, i.e., $f(w) \leq 1$, which in turn demands $w \geq - 2/3$. This implies that the universe was decelerating at the end of the matter-dominated era, i.e., $q_{z_{eq}} \geq 0$.

Thus, collecting results we have
\begin{equation}
\label{zT}
z_{T} = f(w)z_{eq} + [f(w) -1],
\end{equation}
where $z_{eq}$, from  (\ref{correspondence}), is 
\begin{equation}
\label{z eq as a function of Omega Matter and w}
z_{eq} = \left(\frac{\Omega_{M}}{1 - \Omega_{M}}\right)^{1/3w} - 1,
\end{equation}
and the values of $w$ compatible with causality are
\begin{equation}
\label{causality condition}
- 2/3 < w < - 1/3, \end{equation}
or equivalently $1/3 < n < 2/3$. For other values the transition is not causal. 

\paragraph{Vacuum energy:}

As an example, let us consider  the models where the dark energy is the cosmological constant (i.e., $w = -1$). From (\ref{zT}) and (\ref{z eq as a function of Omega Matter and w}) we find 
\begin{equation}
\label{zT and zeq for vacuum energy}
z_{T} - z_{eq} = \left(\frac{1 - \Omega_{M}}{\Omega_{M}}\right)^{1/3}(2^{1/3} - 1) > 0.
\end{equation}
If we take  $\Omega_{M} = 0.3$, then
\begin{equation}
\label{numbers for the cosmological constant}
q_{z_{eq}} = -1/2, \;\;\;z_{eq} \approx 0.326, \;\;\;t{eq} \approx 0.702/\bar{H}, \;\;\;z_{T} \approx 0.671, \;\;\;t_{T} \approx 0.534/\bar{H}.
\end{equation}
 According to this, in models with vacuum energy (or Cardassian expansion with $n = 0$), the universe starts accelerating {\it long before} ($t_{T} \approx 0.534/\bar{H}$) the matter-dominated era comes to an end ($t{eq} \approx 0.702/\bar{H}$). 

\subsection{Cosmological parameters and causality}

Let us find the condition for causality in terms of $\Omega_{M}$ and $\bar{q}$. From  (\ref{q today in terms of Omega and w}) we obtain

\begin{equation}
\label{w in terms of zeq and OmegaM}
w = \frac{(2\bar{q} - 1)}{3(1 - \Omega_{M})}.
\end{equation}
Substituting this expression into  (\ref{f(w)}), (\ref{zT}) and  (\ref{z eq as a function of Omega Matter and w}), we find
\begin{eqnarray}
z_{eq} &=& \left(\frac{1 - \Omega_{M}}{\Omega_{M}}\right)^{(1 - \Omega_{M})/(1 - 2\bar{q})} - 1,\nonumber \\
z_{T} &=& \left(\frac{\Omega_{M} - 2\bar{q}}{\Omega_{M}}\right)^{(1 - \Omega_{M})/(1 - 2\bar{q})} - 1.
\end{eqnarray}
Now, causality condition $z_{eq} > z_{T}$, requires
\begin{equation}
(\Omega_{M} - \bar{q}) < \frac{1}{2}.
\end{equation}
Which, by virtue of (\ref{acc at the end of matter dominated phase}) implies that the universe is decelerating at the end of the matter dominated era.

\subsection{Fitting observational data for $q$} For $\Omega = 0.3$ causality requires $\bar{q} \geq - 0.2$. However, all estimates of the deceleration parameter today; direct, indirect and theoretical,   predict $\bar{q} <  - 0.2$. 

Direct determination  of $q$, from high red-shift supernovae, has recently been provided by John \cite{Moncy} and Daly and Djorgovski \cite{DD1}. John obtains $\bar{q} \approx - 0.77$ in a model-independent cosmographic evaluation of SNe data, without reference to the specific cosmic inventory of energy densities. Daly and Djorgovski obtain $\bar{q} = - 0.35 \pm 0.15$ directly from combinations of the first and second derivatives of the coordinate distance and observational data, without invoking any theory of gravity\footnote{The difference between results in these two papers, which are model-independent,  is mainly due to technical reasons. Namely,   
John expands the scale  factor about zero redshift into a polynomial of order $5$, while  Daly and Djorgovski expand about a
redshift that systematically increases. However, if we add  a fractional error bar to the average value of $\bar{q}  = -0.77$ obtained by John,
we find $\bar{q} = -0.77 \pm 0.33$, which is easily within one $\sigma$ of the
result published by Daly and Djorgovski.}. 

Indirect evidence about $q$ can be obtained from studies of the equation of state for dark energy. Current observations indicate \cite{Riess}-\cite{Tonry}, \cite{Alam}-\cite{Corasaniti} 
\begin{equation}
\label{current w}
- 1.48 < w < - 0.72,
\end{equation}
or equivalently $- 1 < \bar{q} < - 0.26$, which is outside of the range permitted by causality. 

Theoretical models of quintessence, in the framework of  flat FRW cosmologies, like  tracker fields \cite{Steinhard}, indicate $w \approx - 0.7 $. Similarly, in braneworld theories, where our universe is embedded in a bulk with more than four dimensions, the predicted values for $\bar{q}$ are in the interval$(- 0.50, \;\;- 0.32)$ \cite{JPdeL1}-\cite{NewAccExp}.

\section{Conclusion}
 
From the above discussion it follows that the most probable option is $(\Omega_{M} - \bar{q}) > 1/2$. Consequently, in these models the transition from deceleration to acceleration occurs before the matter-dominated epoch comes to an end, i.e., $t_{eq} > t_{T}$. Which contradicts the  most basic of all the principles of physics, namely, the causality principle. A clear example of this is provided by the cosmological constant (\ref{numbers for the cosmological constant}). We advise in-depth  theoretical work and a reexamination of observational basics.

\medskip

\paragraph{Acknowledgements:} I would like to thank Ruth A. Daly and Moncy John for explaining me some details about their respective papers.

\end{document}